# A geometric approach to separate the effects of magnetic susceptibility and chemical shift/exchange in a phantom with isotropic magnetic susceptibility

Hyunsung Eun[1*], Hwihun Jeong[1*], Jingu Lee[1], Hyeong-geol Shin[1], and Jongho Lee[1]

* Hyunsung Eun and Hwihun Jeong contributed equally to this work.

**Author affiliations:**

[1]Laboratory for Imaging Science and Technology, Department of Electrical and Computer Engineering, Seoul National University, Seoul, Korea

**Correspondence**

Jongho Lee, Ph. D

Department of Electrical and Computer Engineering, Seoul National University

Building 301, Room 1008, 1 Gwanak-ro, Gwanak-gu, Seoul, Korea

Tel: 82-2-880-7310

E-mail: jonghoyi@snu.ac.kr

**Word count**: 3,153

# ABSTRACT

**Purpose:** To separate the effects of magnetic susceptibility and chemical shift/exchange in a phantom with isotropic magnetic susceptibility. To generate a chemical shift/exchange-corrected quantitative susceptibility mapping (QSM) result.

**Theory and Methods:** Magnetic susceptibility and chemical shift/exchange are the properties of a material. Both are known to induce the resonance frequency shift in MRI. In current QSM, the susceptibility is reconstructed from the frequency shift, ignoring the contribution of the chemical shift/exchange. In this work, a simple geometric approach, which averages the frequency shift maps from three orthogonal $B_0$ directions to generate a chemical shift/exchange map, is developed using the fact that the average nullifies the (isotropic) susceptibility effects. The resulting chemical shift/exchange map is subtracted from the total frequency shift, producing a frequency shift map solely from susceptibility. Finally, this frequency shift map is reconstructed to a susceptibility map using a QSM algorithm. The proposed method is validated in numerical simulations and applied to phantom experiments with olive oil, bovine serum albumin, ferritin, and iron oxide solutions.

**Results:** Both simulations and experiments confirm that the method successfully separates the contributions of the susceptibility and chemical shift/exchange, reporting the susceptibility and chemical shift/exchange of olive oil (susceptibility: 0.62 ppm, chemical shift: -3.60 ppm), bovine serum albumin (susceptibility: -0.059 ppm, chemical shift: 0.008 ppm), ferritin (susceptibility: 0.125 ppm, chemical shift: -0.005 ppm), and iron oxide (susceptibility: 0.30 ppm, chemical shift: -0.039 ppm) solutions.

**Conclusion:** The proposed method successfully separates the susceptibility and chemical shift/exchange in phantoms with isotropic magnetic susceptibility.

## 1. INTRODUCTION

With the development of ultra-high field MRI systems, phase shift or resonance frequency shift has become a vital contrast for high-resolution anatomy.[1-3] The origins of the resonance frequency shift have been attributed to the effect of magnetic susceptibility,[4-6] chemical shift,[7] chemical exchange,[8-11] and complex tissue microstructure.[12-14] Among them, the magnetic susceptibility effect has been suggested as a primary contributor to the resonance frequency shift. Under this assumption, quantitative susceptibility mapping (QSM) has been proposed by converting a frequency shift image to a susceptibility map.[15-19] The method has been demonstrated to generate great details of anatomical structures and has been applied for various clinical studies.[20-24] However, the accuracy of QSM degrades when other sources of frequency shift exist. For example, in abdominal QSM, the chemical shift from fat introduces substantial artifacts in QSM images if it is reconstructed without consideration of the chemical shift of fat.[25,26] Efforts were made to design models that separated the susceptibility and non-susceptibility sources but they required specific conditions to hold (e.g. a piece-wise constant assumption).[27,28]

The magnetic susceptibility and chemical shift/exchange are the properties of a material (e.g., vegetable oil has the magnetic susceptibility of 0.65 ppm and chemical shift of -3.46 ppm;[28] bovine serum albumin solution (100 mg/ml) has the magnetic susceptibility of -0.068 ppm and chemical exchange of 0.008 ppm [29]; see Discussion). Hence, understanding the contribution of the two sources is important not only for the reliable reconstruction of QSM, but also for the separate measurements of the two sources. In a few previous studies,[8-10] efforts have been made to separate the effects of the magnetic susceptibility and chemical exchange by mixing a material of interest with a reference chemical (e.g., dioxane). However, a more recent study has suggested a non-negligible level of interactions between the reference chemical and the material under investigation.[30] As a result, further research is necessary to determine the validity of the reference chemical-based measurements.

In this study, we present a novel geometric method that separates the susceptibility and chemical shift/exchange without requiring a reference chemical that is mixed with the sample of interest. A mathematical formation that utilizes multiple $B_0$ orientation data has been developed to achieve the separation. Numerical simulation and phantom experiments were performed to validate the method.

## 2. THEORY

### 2.1. Resonance frequency shift

When the effect of the susceptibility and chemical shift/exchange coexist in an isotropic medium (i.e. no anisotropic susceptibility, no anisotropic microstructure, and no orientation-dependent chemical shift/exchange) with no additional sources for the frequency shift, the total frequency shift can be written as

$$\Delta f(\vec{r}) = f_c(\vec{r}) + f_s(\vec{r}), \qquad [1]$$

where $\Delta f$ is the total resonance frequency shift, $\vec{r}$ is a position vector, $f_c$ is the chemical shift/exchange-induced frequency shift, and $f_s$ is the susceptibility-induced frequency shift. The susceptibility-induced frequency shift has been shown to be modeled as:[31,32]

$$f_s(\vec{r}) = d * \chi(\vec{r}), \qquad [2]$$

$$d(\vec{r}) = \frac{1}{4\pi} \cdot \frac{3\cos^2\theta - 1}{\vec{r}^3}, \qquad [3]$$

where $d$ is a dipole kernel, $\chi$ is susceptibility, and $\theta$ is an angle between the position vector and $B_0$ orientation. Since $d$ is a function of $B_0$ orientation, the susceptibility-induced frequency shift is also dependent on the orientation of $B_0$.[4,33]

### 2.2. Separation of chemical shift/exchange from susceptibility

When $B_0$ is applied to an object along the three orthogonal axes (i.e., x-, y-, or z-axis), the three susceptibility-induced frequency shifts can be written as

$$f_{s,x}(\vec{r}) = d_x(\vec{r}) * \chi(\vec{r}) = \frac{1}{4\pi} \cdot \frac{3\cos^2\theta_x - 1}{\vec{r}^3} * \chi(\vec{r}),$$

$$f_{s,y}(\vec{r}) = d_y(\vec{r}) * \chi(\vec{r}) = \frac{1}{4\pi} \cdot \frac{3\cos^2\theta_y - 1}{\vec{r}^3} * \chi(\vec{r}),$$

$$f_{s,z}(\vec{r}) = d_z(\vec{r}) * \chi(\vec{r}) = \frac{1}{4\pi} \cdot \frac{3\cos^2\theta_z - 1}{\vec{r}^3} * \chi(\vec{r}), \qquad [4]$$

where sub-index x, y, and z indicate the orientation of $B_0$ and $\theta_x$, $\theta_y$, and $\theta_z$ represent the angle between the position vector $\vec{r}$ and the $B_0$ field along the x, y, and z-axes, respectively.

When we sum the three frequency shifts of the orthogonal orientations, it results in a null frequency shift:

$$f_{s,x}(\vec{r}) + f_{s,y}(\vec{r}) + f_{s,z}(\vec{r}) = \left\{ \frac{1}{4\pi} \cdot \frac{3(\cos^2\theta_x + \cos^2\theta_y + \cos^2\theta_z) - 3}{\vec{r}^3} \right\} * \chi(\vec{r}) = 0, \quad [5]$$

because $\cos^2\theta_x + \cos^2\theta_y + \cos^2\theta_z$ equals to 1 for the three orthogonal axes (see Supporting Information Section 1).

Using this property, we can measure the chemical shift/exchange-induced frequency shift by averaging the frequency shifts from the three orthogonal $B_0$ orientations.

$$f_{average}(\vec{r}) = \frac{1}{3}\{f_x(\vec{r}) + f_y(\vec{r}) + f_z(\vec{r})\}$$

$$= \frac{1}{3}[3f_c(\vec{r}) + \{f_{s,x}(\vec{r}) + f_{s,y}(\vec{r}) + f_{s,z}(\vec{r})\}] = f_c(\vec{r}). \quad [6]$$

Then, the susceptibility-induced frequency shift can be calculated by subtracting the chemical shift/exchange-induced frequency shift from the total frequency shift.

$$f_s(\vec{r}) = f(\vec{r}) - f_c(\vec{r}). \quad [7]$$

From this susceptibility-induced frequency shift, one can measure the susceptibility by using a QSM algorithm.

This approach of separating the susceptibility and chemical shift/exchange requires data acquisition of three different $B_0$ orientations that are orthogonal to each other. If an object has geometric symmetry, however, the number of scans can be reduced. For example, a cylindrical phantom has the geometric symmetry in the two short axes and, therefore, only two acquisitions, one with $B_0$ along the long axis of the cylinder and the other with $B_0$ along one of the short axes, are necessary. The missing frequency shift map can be generated by the 90° rotation of the frequency shift map acquired with $B_0$ along the short axis. Similarly, a spherical phantom can be scanned once to apply our method. The idea can be generalized for arbitrary $B_0$ orientations using a least-squares minimization method (see Supporting Information Section 2).

## 3. METHODS

### 3.1. Numerical simulation

A numerical simulation was designed to demonstrate that the summation of the three susceptibility-induced frequency shifts results in a null field (Eq. 5). Four numerical phantoms in the shape of heart, cylinder, sphere, and brain[34] were constructed in a $256^3$-voxel grid each. For the heart, cylinder, and sphere phantoms, the susceptibility of 0.1 ppm was assigned. For the brain phantom, the following susceptibility values were assigned to subregions: caudate nucleus = 0.04 ppm, putamen = 0.07 ppm, globus pallidus = 0.12 ppm, gray matter = 0.02 ppm, and white matter = -0.02 ppm. The background had zero susceptibility. No chemical shift/exchange was assumed. Then, the susceptibility-induced frequency shifts of the three orthogonal $B_0$ orientations were generated using a Fourier-based frequency shift calculation method. After that the three susceptibility-induced frequency shifts were summed to demonstrate that the result was a null field.

Another numerical simulation was performed to illustrate the process of the proposed method. A cylindrical phantom of the 30-voxel radius was embedded at the center of a 256×256×40-voxel grid. The material inside the phantom was assumed to be fat with a chemical shift of -3.5 ppm[35] and susceptibility of 0.65 ppm[28]. The background was assumed to have zero susceptibility and chemical shift. A susceptibility-induced frequency shift map was calculated for each of the three orthogonal $B_0$ orientations using the Fourier-based method. Then a chemical shift-induced frequency shift map, which was confined to the cylinder with a -3.5 ppm frequency shift, was generated. The two maps were summed to yield a total frequency shift map for each $B_0$ orientation. Our method of separating the susceptibility and chemical shift sources was applied to the three total frequency shift maps. As the first step, the three total frequency shift maps were averaged to produce a chemical shift-induced frequency shift map (Eq. 6). Then, this map was subtracted from each of the total frequency shift maps, providing the susceptibility-induced frequency shift maps (Eq. 7). From these frequency shift maps, susceptibility maps of each orientation were reconstructed by a QSM method[36] with a regularization factor of 10. For comparison, susceptibility maps were reconstructed from the total frequency shifts, which contained the chemical shift effect, using the same QSM method.

### 3.2. Phantom experiment

Olive oil (O1514, Sigma-Aldrich, St.Louis, MO), bovine serum albumin (BSA) solution (100 mg/ml; A7906, Sigma-Aldrich), ferritin solution (0.43 mg/ml; F4503, Sigma-Aldrich), and iron oxide solution ($2.5 \cdot 10^{-3}$ mg/ml; PMC1N, Bang's Laboratories Inc., Fishers, IL) were used to test the proposed method. Each solution filled out two identical plastic cylinders (inner diameter = 28 mm, outer diameter = 30 mm, and height = 110 mm). Then, the two cylinders were positioned in a large container such that one cylinder was parallel to $B_0$ and the other was perpendicular to $B_0$. The container was filled with distilled water. In a separate scan, a sphere phantom (diameter = 40 mm) filled with the olive oil was also constructed and tested.

All scans were conducted on a 3T MRI scanner (SIEMENS Tim Trio, Erlangen, Germany) with a 12-channel array coil. The scan started with a three-plane localizer. Then shimming was performed to reduce field inhomogeneity. For the main scan of the BSA, ferritin, and iron oxide solutions, data were acquired at room temperature using a 3D multi-echo gradient echo sequence with the following parameters: FOV = 192×192×64 mm$^3$, voxel size = 2×2×2 mm$^3$, TR = 64 ms, TE = 5 to 23.75 ms with an echo spacing of 3.75 ms, number of echoes = 6, bandwidth = 330 Hz/pixel, flip angle = 12°, GRAPPA factor = 2, monopolar readout gradient, and total acquisition time = 2.03 min. For the olive oil, the readout gradient was modified from monopolar to bipolar in order to achieve shorter echo spacing, because of the large chemical shift of the olive oil when compared to the others. The scan parameters were also modified to TR = 40 ms, TE = 1.5 to 9.78 ms with an echo spacing of 0.92 ms, number of echoes = 10, bandwidth = 2000 Hz/pixel, and total acquisition time = 1.6 min. To compensate for the artifacts in the bipolar gradient, the sequence was repeated with the opposite gradient polarity.[37] After scanning each solution, the two cylinders (or a sphere) were replaced by cylinders (or a sphere) with distilled water. Then the scan was repeated to acquire a reference for a background field.[38]

For data processing, the k-space data of each solution were reconstructed to multi-echo complex images via GRAPPA reconstruction[39] and multi-channel image combination[40]. For the olive oil data, the two same TE complex images of the opposite gradient polarity were averaged to generate a complex image.[37] The background field was removed by dividing the complex image of the main scan by that of the reference. Then, a frequency shift map was derived using a linear least-squares fitting from the multi-echo data. For the cylindrical phantoms, the two frequency shift maps, one parallel and the other perpendicular to $B_0$, were manually registered. After that, the third orientation frequency shift map was generated by a 90° rotation of the frequency shift map whose cylinder was perpendicular to $B_0$. The resulting map was registered to the other maps. For the spherical phantom, the frequency shift map was rotated by 90° in two orthogonal orientations. Then the two rotated maps were registered to the original map. After that our method was applied to separate the susceptibility-induced frequency shift and chemical shift/exchange-induced frequency shift. From the susceptibility-induced frequency shift maps, susceptibility maps were reconstructed using the QSM method with two different regularization factors (10 and 100). During the QSM reconstruction, voxels that contained the cylinder wall had low signal to noise ratios (SNR) and were excluded to avoid partial volume-induced errors. For comparison, susceptibility maps were also reconstructed from the total frequency shift maps. To quantify the results, a region of interest (ROI) was chosen at the central eight slices of the cylinder. Then the mean and standard deviation of the susceptibility and chemical shift/exchange were measured in the ROI.

All data processing was performed using MATLAB (MATLAB 2014b, MathWorks Inc., Natick, MA).

## 4. RESULTS

The simulation results of the four numerical phantoms in the shapes of the heart, cylinder, sphere, and brain confirm that the sum of the three orthogonal frequency shift maps generates a null frequency shift in all phantoms (Supporting Information Section 3), demonstrating the validity of Eq. 5 in Theory.

The process of separating the susceptibility and chemical shift/exchange is shown in Figure 1 for the numerical fat phantom. The total frequency shift maps of three orthogonal $B_0$ orientations are illustrated in Figure 1a. When the three frequency shift maps are averaged to generate a chemical shift map (Fig. 1b), the frequency shift measured inside the cylinder is -3.5 ppm, which is the same as the assigned chemical shift. This map is subtracted from each of the total frequency shift maps, generating the susceptibility-induced frequency shift maps, as shown in Figure 1c. When these maps are reconstructed for QSM, they result in consistent susceptibility estimations for all $B_0$ orientations (Fig. 1e; left cylinder: 0.65 ± 0.01 ppm, middle cylinder: 0.65 ± 0.01 ppm, right cylinder: 0.65 ± 0.00 ppm) that are equal to the assigned susceptibility value (0.65 ppm) regardless of the $B_0$ orientations. On the other hand, when susceptibility maps are reconstructed from the total frequency shifts in Figure 1a, the susceptibility values show significant variations (Fig. 1d; left cylinder: 4.16 ± 0.60 ppm, middle cylinder: 4.16 ± 0.60 ppm, right cylinder: -8.97 ± 0.01 ppm). These results demonstrate that the chemical shift/exchange-induced frequency shift produces susceptibility errors and streaking artifacts.

The experimental results using the cylindrical olive oil phantom are displayed in Figure 2. The same trend as in the numerical fat phantom is observed. The frequency shift maps from the three orthogonal $B_0$ show orientation dependency (Fig. 2a). When the three maps are averaged, the chemical shift-induced frequency shift map reveals -3.60 ± 0.01 ppm (Fig. 2b). This map shows no blooming artifacts, confirming that the chemical shift effect is localized to the source. After removing the chemical shift effect, the susceptibility-induced frequency shift maps (Fig. 2c) are reconstructed to generate the QSM maps (Fig. 2e). The results report consistent susceptibility for all orientations (left cylinder: 0.63 ± 0.02 ppm, middle cylinder: 0.61 ± 0.02 ppm, right cylinder: 0.63 ± 0.01 ppm). The measurements show robustness to the regularization factors (Supporting Information, Table S1). When susceptibility maps are reconstructed from the total frequency shift maps that include the chemical shift effect, they result in large susceptibility variations (left cylinder: 11.4 ± 0.8 ppm, middle cylinder: 10.6 ± 1.1 ppm, right cylinder: -10.2 ± 0.2 ppm) and streaking artifacts (Fig. 2d). The results of the olive oil using a spherical phantom report the same trend and are summarized in Supporting Information Figure S2.

The susceptibility and chemical exchange measurements of the BSA, ferritin, and iron oxide solutions are listed in Table 1. In all solutions, the susceptibility measurements show more consistent results in the proposed method than in the conventional QSM reconstruction.

## 5. DISCUSSION

In this paper, a novel method that separates the susceptibility and chemical shift/exchange in a phantom without anisotropy susceptibility or microstructure is presented. The method yields quantitative maps of magnetic susceptibility and chemical shift/exchange and is tested in numerical simulations and phantom experiments.

When the susceptibility and chemical shift/exchange measurements from our method were compared to literature values, they showed a good correspondence. The measurements of the BSA solution (susceptibility: -0.059 $\pm$ 0.002 ppm, chemical exchange: 0.008 $\pm$ 0.001 ppm) closely matched to the literature values (susceptibility: -0.068 ppm, chemical exchange: 0.008 ppm; these values were scaled from Reference 29 for concentration and Lorentz sphere correction). The measurements of the olive oil (susceptibility: 0.62 $\pm$ 0.01 ppm, chemical shift: -3.60 $\pm$ 0.01 ppm) were also well in agreement with the literature values (susceptibility: 0.65 ppm and chemical shift: -3.46 ppm from vegetable oil measured in Reference 28; susceptibility: 0.75 ppm from oleic acid in Reference 41).

In previous studies,[8-11] the quantification of chemical shift/exchange of a solution was performed using a reference chemical (e.g., dioxane), assuming no interaction between the reference and solution. In a more recent study, however, a non-negligible amount of interaction was reported[30], undermining the results of previous studies[8-11]. In other studies, algorithms were proposed to separate susceptibility and non-susceptibility sources, assuming a piece-wise constant condition.[27,28] Our approach, on the other hand, avoids potential complications from the reference chemical while allowing voxel-wise image reconstruction.

The proposed method requires three scans of orthogonal $B_0$ orientations when an object has an arbitrary shape. The number of scans can be reduced in an object with symmetry (e.g., sphere or cylinder). For example, a spherical phantom can be scanned just once. However, practical issues such as positioning the sphere, removing air bubbles, and performing the reference scan were difficult in the spherical phantom. As a result, it was used once to demonstrate the feasibility. Instead, cylindrical phantoms, which require two scans of orthogonal $B_0$ orientations, are easier to handle. The two phantoms resulted in consistent measurements as demonstrated in Figure 2 and Supporting Information Figure S2.

When the alignment of the three orthogonally oriented dataset is not accurate, it may introduce errors in the estimations. This effect is analyzed in Supporting Information Section 6. The results suggest that the measurements are robust (less than 0.5% errors) for a small misalignment (< 5°). Addtionally, the effects of the noise amplification of the proposed method is analyzed in Supporting Information Section 7.

In our study, a reference scan was acquired in order to eliminate the background field accurately. Alternatively, one may apply a background field removal method to reduce the scan time from the reference scan.[42]

Since the olive oil phantom has a large chemical shift-induced frequency shift (-3.58 ppm; 458.24 Hz in 3T), a shorter echo spacing time (<1.09 ms) was required for the GRE sequence to reconstruct the frequency shift maps from the phase maps. To reduce the echo spacing time, we modified the readout gradient of the sequence from monopolar to bipolar. Then, the two datasets with the opposite readout gradient were combined to avoid the phase modulations.[37] This modified GRE sequence with the echo spacing of 0.92 ms allowed us to reconstruct the frequency shift maps of olive oil phantom without any artifacts.

In the phantom experiments, we experienced a $B_0$ drift, which may induce measurement errors.[43] To minimize the $B_0$ drift-induced errors, we stabilized the $B_0$ field through a sufficient amount of dummy scans before the main scan. The scan time was also reduced by using parallel imaging.[39]

Applying our method to in-vivo human brain is challenging because of the three orthogonal orientation scans required. Furthermore, the brain has additional sources of frequency shift, such as tissue compartmentalization[13] and susceptibility anisotropy.[6] These contributions have $B_0$ orientation dependency and are not considered in our model. Additionally, a study has suggested that chemical exchange may have $B_0$ orientation dependency in in-vivo.[44] All of these will complicate our results when the method is applied directly to in-vivo. As a result, further improvement in methodology is necessary and this work will provide a stepping stone toward future research.

## 6. CONCLUSION

In this study, we developed a geometric approach for separating the susceptibility and chemical shift/exchange in a phantom without anisotropic susceptibility or microstructure. The proposed method yields accurate susceptibility and chemical shift/exchange measurements in the numerical simulation and phantom experiments. When susceptibility maps are reconstructed from the total frequency shift maps that include the chemical shift/exchange effect, they reveal large susceptibility variations and streaking artifacts depending on the $B_0$ orientation. On the other hand, the reconstruction results from the susceptibility-induced frequency shift maps demonstrate correct susceptibility measurements with no artifacts. The proposed method is useful not only in measuring magnetic susceptibility and chemical shift/exchange but also in improving QSM reconstruction algorithms.

**ACKNOWLEDGEMENTS**

This work was supported by the National Research Foundation of Korea grant funded by the Korea government (MSIT; NRF-2018R1A2B3008445), Creative-Pioneering Researchers Program through Seoul National University, and Institute of Engineering Research at Seoul National University.

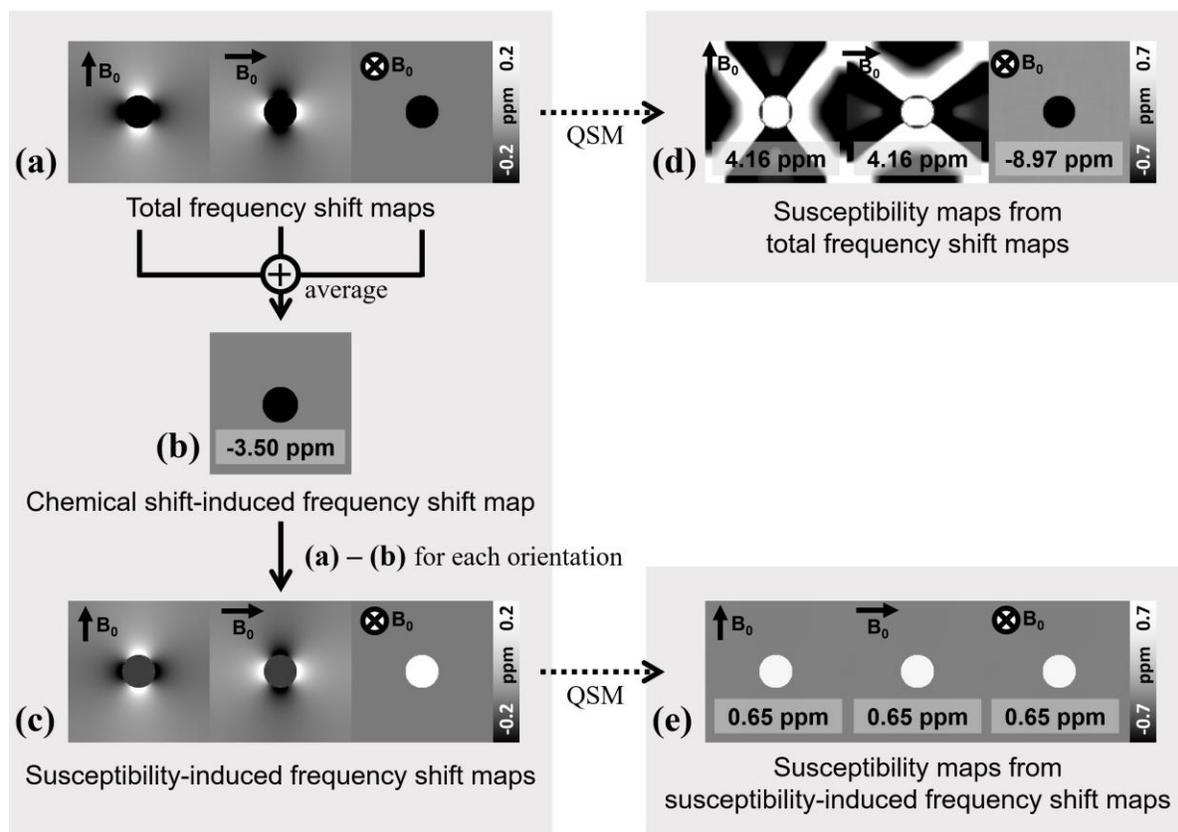

**Figure 1.** Separation of the susceptibility and chemical shift in the numerical fat phantom. The three total frequency shift maps generated from the three orthogonal $B_0$ orientations include the contributions of both susceptibility and chemical shift (a). The average of the three total frequency shift maps produces a chemical shift-induced frequency shift map (b). The susceptibility-induced frequency shift maps are generated by subtracting the chemical shift-induced frequency shift from the total frequency shift maps (c). The susceptibility maps reconstructed from the total frequency shift maps (d), and the susceptibility maps from the susceptibility-induced frequency shift maps (e) reveal significant differences with the latter reporting correct susceptibility estimations in all orientations.

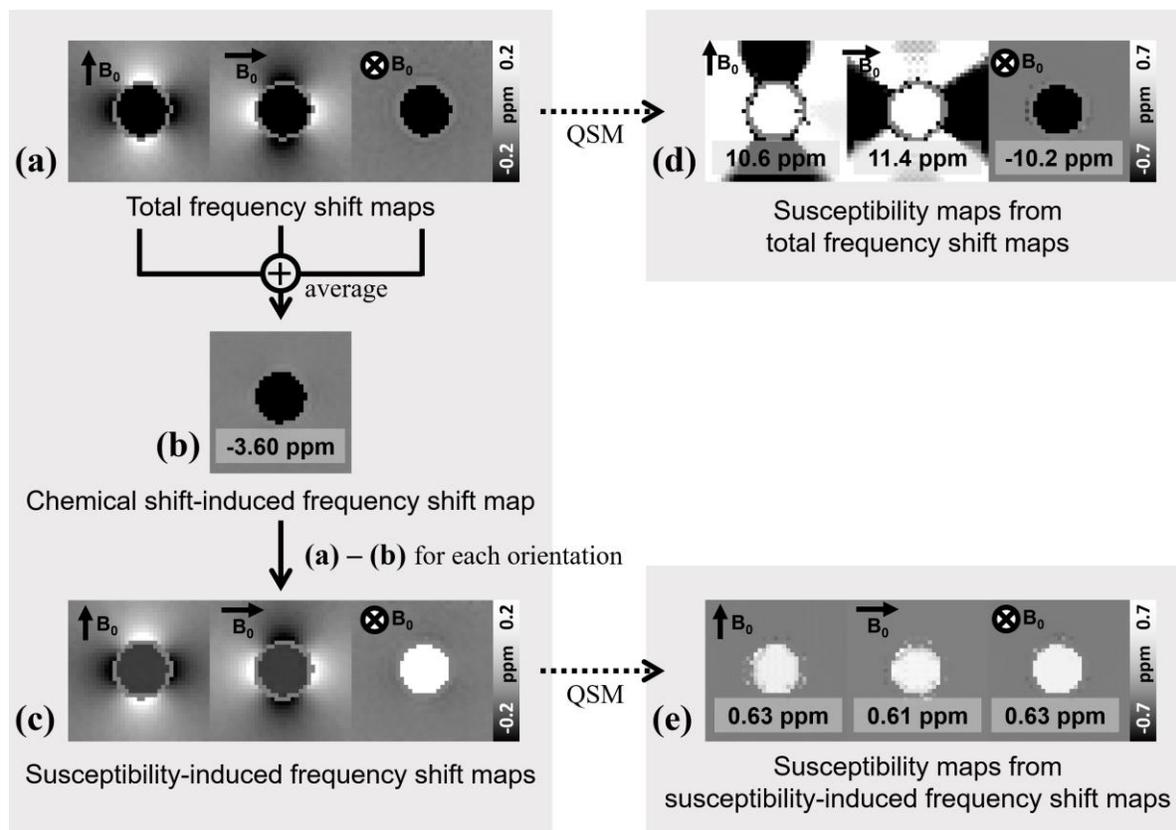

**Figure 2.** Separation of the susceptibility and chemical shift in the cylindrical phantom with olive oil. The three total frequency shift maps generated from the three orthogonal $B_0$ orientations contain the effect of both susceptibility and chemical shift (a). The average of the three total frequency shift maps produces the chemical shift-induced frequency shift map (b). The susceptibility-induced frequency shift maps are generated by subtracting the chemical shift-induced frequency shift from the total frequency shift maps (c). The susceptibility maps reconstructed from the total frequency shift maps (d) and the susceptibility maps from the susceptibility-induced frequency shift maps (e) demonstrate significant differences with the latter reporting consistent and correct susceptibility measurements in all orientations.

# Table 1

Susceptibility (χ) and Chemical Shift/Exchange (Ξ) Measurements for the Olive Oil, BSA, Ferritin, and Iron Oxide Solutions.

| | Ξ, ppm | χ, ppm | | B₀ |
|---|---|---|---|---|
| | | Susceptibility measurements using proposed method | Susceptibility measurements from total frequency shift | |
| Olive oil | -3.60 ± 0.01 | 0.63 ± 0.02 | 11.4 ± 0.8 |  |
| | | 0.61 ± 0.02 | 10.6 ± 1.1 |  |
| | | 0.63 ± 0.01 | -10.2 ± 0.2 |  |
| | -3.58 ± 0.01 | 0.66 ± 0.01 | -5.07 ± 0.63 |  |
| | | 0.66 ± 0.01 | -4.02 ± 1.08 |  |
| | | 0.62 ± 0.01 | -4.23 ± 0.07 |  |
| BSA | 0.008 ± 0.001 | -0.057 ± 0.001 | -0.082 ± 0.001 |  |
| | | -0.058 ± 0.002 | -0.080 ± 0.002 |  |
| | | -0.061 ± 0.001 | -0.038 ± 0.001 |  |
| Ferritin | -0.005 ± 0.001 | 0.126 ± 0.005 | 0.137 ± 0.006 |  |
| | | 0.124 ± 0.006 | 0.136 ± 0.008 |  |
| | | 0.126 ± 0.001 | 0.109 ± 0.001 |  |
| Iron oxide | -0.039 ± 0.003 | 0.28 ± 0.01 | 0.43 ± 0.01 |  |
| | | 0.31 ± 0.01 | 0.45 ± 0.01 |  |
| | | 0.32 ± 0.01 | 0.21 ± 0.01 |  |

# Supporting Information

### Section 1. Three cosine terms in Equation 5 sums zero

For the position vector ($\vec{r}$), the ratio of each component of the vector to the length of the vector can be expressed as the following cosine terms:

$$\cos\theta_x = \frac{\|x\|}{\|\vec{r}\|},$$

$$\cos\theta_y = \frac{\|y\|}{\|\vec{r}\|},$$

$$\cos\theta_z = \frac{\|z\|}{\|\vec{r}\|},$$

where $\|\vec{r}\|^2 = x^2 + y^2 + z^2$.

Therefore, $\cos^2\theta_x + \cos^2\theta_y + \cos^2\theta_z = \frac{x^2+y^2+z^2}{\|\vec{r}\|^2} = 1$, confirming Equation 5 to be zero.

## Section 2. Least-squares solution for arbitrary B₀ orientation data

In this section, we generalize our method for non-orthogonal B₀ orientation data.

The Fourier transform of the total resonance frequency shift with $N$ different B₀ orientations can be modeled as follows:

$$\Delta F_i(\vec{k}) = F_c(\vec{k}) + F_{s,i}(\vec{k}) = F_c(\vec{k}) + D_i(\vec{k})X(\vec{k}), \quad [S1]$$

$$D_i(\vec{k}) = \frac{1}{3} - \frac{(k_z \cos\theta_i + k_y \sin\theta_i \cos\phi_i + k_x \sin\theta_i \sin\phi_i)^2}{k_x^2 + k_y^2 + k_z^2}, \quad [S2]$$

where $\Delta F$ is the total frequency shift in k-space, $F_s$ is the susceptibility-induced frequency shift in k-space, $F_c$ is the chemical shift/exchange-induced frequency shift in k-space, $D$ is a dipole kernel in k-space, and $X$ is susceptibility in k-space. The sub-index $i$ represents a B₀ orientation ($i$ = 1, 2, …, $N$), and $\theta_i$ and $\phi_i$ are the angles of the B₀ orientation in the spherical coordinate at the $i^{th}$ orientation.

Given a set of $N$ measurements, the frequency shifts can be reformulated as follows:

$$\begin{bmatrix} \Delta F_1(\vec{k}) \\ \Delta F_2(\vec{k}) \\ \vdots \\ \Delta F_N(\vec{k}) \end{bmatrix} = \begin{bmatrix} 1 & D_1(\vec{k}) \\ 1 & D_2(\vec{k}) \\ \vdots & \vdots \\ 1 & D_N(\vec{k}) \end{bmatrix} \begin{bmatrix} F_c(\vec{k}) \\ X(\vec{k}) \end{bmatrix} = D \begin{bmatrix} F_c(\vec{k}) \\ X(\vec{k}) \end{bmatrix}. \quad [S3]$$

If a sufficient number of data with different angles are acquired, the inverse problem in Equation S3 can be well-conditioned. Then, both chemical shift/exchange and susceptibility maps can be reconstructed using the least-squares estimation as shown below.

$$\begin{bmatrix} F_c(\vec{k}) \\ X(\vec{k}) \end{bmatrix} = (D^T D)^{-1} D^T \begin{bmatrix} \Delta F_1(\vec{k}) \\ \Delta F_2(\vec{k}) \\ \vdots \\ \Delta F_N(\vec{k}) \end{bmatrix}. \quad [S4]$$

**Section 3. Numerical phantoms validating Theory**

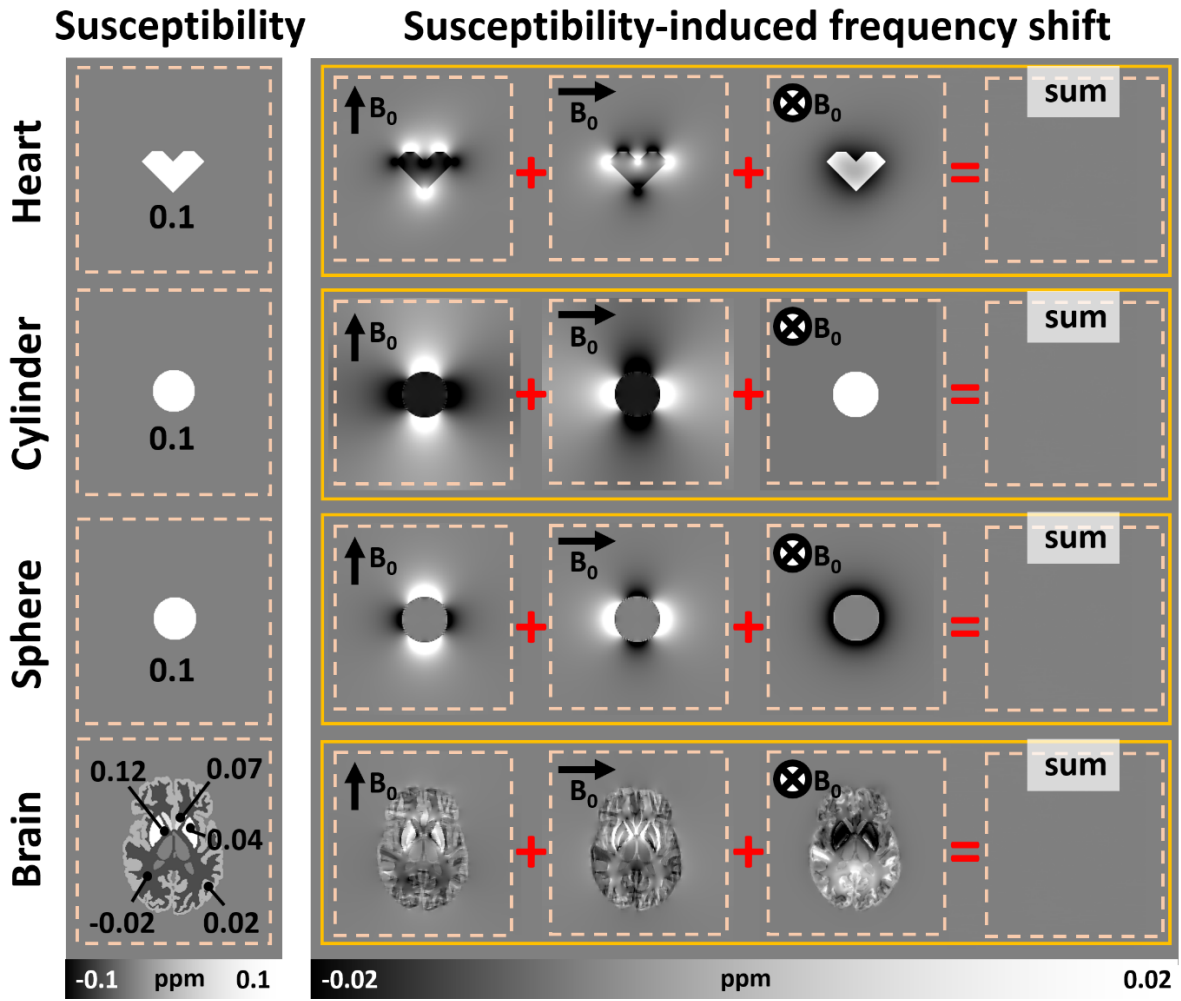

**Figure S1.** Susceptibility-induced frequency shifts in the heart-, cylinder-, sphere- and brain-shaped numerical phantoms. The frequency shift maps demonstrate $B_0$ orientation dependency. When the frequency shift maps of the three orthogonal $B_0$ orientations are summed, a null field is generated. The susceptibility and frequency shifts are expressed in parts per million.

# Section 4. Effects of two different regularization factors

**Table S1.** Susceptibility measurements with two different regularization factors ($\lambda = 10$ and $\lambda = 100$).

| | $B_0$ | Susceptibility measurements in ppm | |
|---|---|---|---|
| | | $\lambda = 10$ | $\lambda = 100$ |
| Olive oil |  | 0.63 ± 0.02 | 0.63 ± 0.02 |
| |  | 0.61 ± 0.02 | 0.63 ± 0.02 |
| |  | 0.63 ± 0.01 | 0.63 ± 0.01 |
| BSA |  | -0.057 ± 0.001 | -0.059 ± 0.001 |
| |  | -0.058 ± 0.002 | -0.059 ± 0.001 |
| |  | -0.061 ± 0.001 | -0.061 ± 0.001 |
| Ferritin |  | 0.126 ± 0.005 | 0.130 ± 0.004 |
| |  | 0.124 ± 0.006 | 0.131 ± 0.004 |
| |  | 0.126 ± 0.001 | 0.126 ± 0.001 |
| Iron oxide |  | 0.28 ± 0.01 | 0.29 ± 0.01 |
| |  | 0.31 ± 0.01 | 0.32 ± 0.01 |
| |  | 0.32 ± 0.01 | 0.32 ± 0.01 |

## Section 5. Experimental results of the olive oil in the spherical phantom

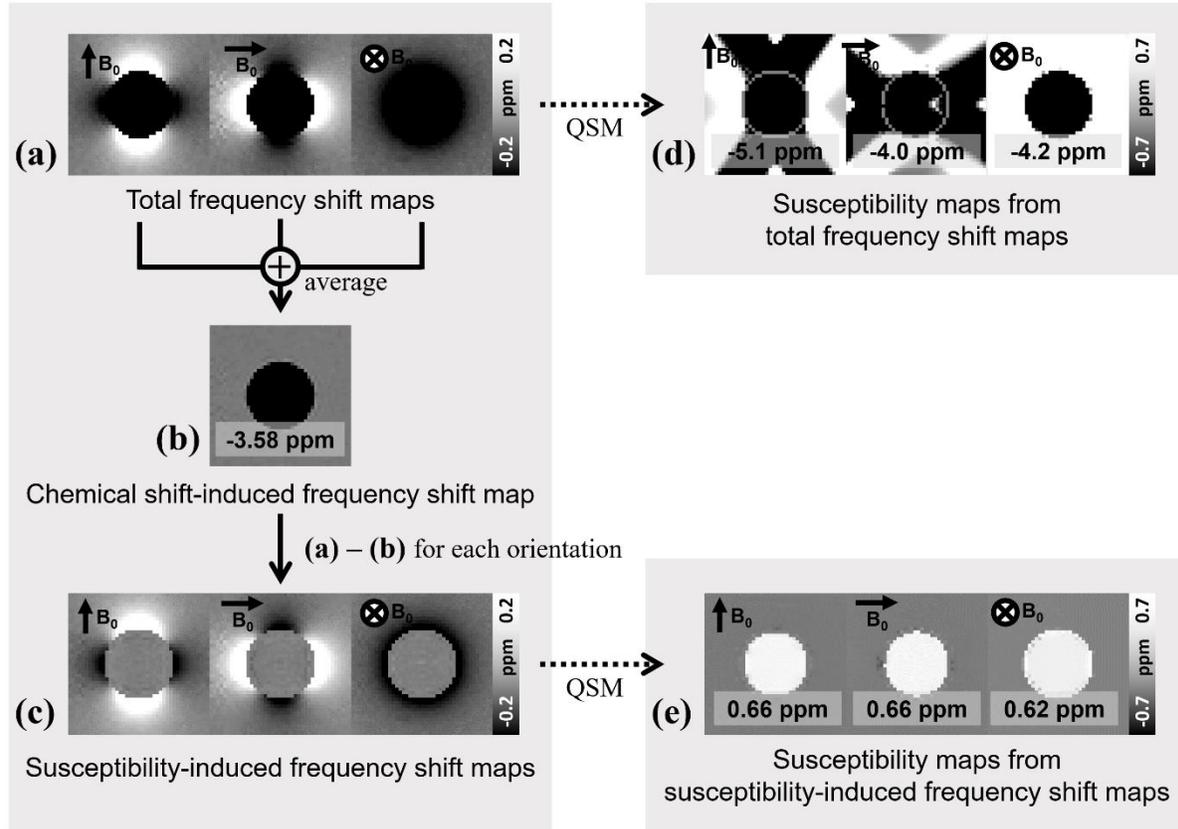

**Figure S2.** Experimental results of the olive oil in the spherical phantom. The chemical shift (b, -3.58 ± 0.01 ppm) and susceptibility (e, left sphere: 0.66 ± 0.01 ppm, middle sphere: 0.66 ± 0.01 ppm, right sphere: 0.62 ± 0.01 ppm) measurements are close to those of the olive oil in the cylindrical phantom. When the susceptibility maps are reconstructed from the total frequency shift maps with the chemical shift effect (b), significant susceptibility variations (left sphere: -5.07 ± 0.63 ppm, middle sphere: -4.02 ± 1.08 ppm, right sphere: -4.23 ± 0.07 ppm) and streaking artifacts are observed (d). As compared to the cylindrical phantom, the spherical phantom requires only one scan for the proposed method. However, the spherical phantom is not practical in some ways, and therefore, the cylindrical phantoms are used for the other solutions.

**Section 6. Effects of misalignment**

Here, we design a numerical simulation to estimate the effects of misalignment of the cylinders in the experiment. A cylindrical phantom of the 30-voxel radius is embedded at the center of a 256×256×256-voxel grid. The material inside the phantom is assumed to have a chemical shift of -3.5 ppm and susceptibility of 0.65 ppm with respect to the zero susceptibility and zero chemical shift of the background. One of the three cylinders is tilted by an angle $\theta$ centered at the middle grid point while the other two are perfectly aligned. Then, the three frequency shift maps are generated as described in the simulation. During the registration of the three frequency shift maps, the tilted angle was corrected because identifying and correcting for the misalignment in the acquired MR images is easy. Finally, the chemical shift and susceptibility maps are reconstructed using the proposed method. The percentage errors of the reconstructed values are calculated for a range of $\theta$ (from 0° to 10°) for each axis.

The results are summarized in Figure S3. For both susceptibility and chemical shift, the misalignment effects are small, reporting less than 0.03% error for chemical shift and 0.5% for susceptibility when the misalignment error is less than 5°. The mean angular error in our experiment was $0.17 \pm 0.48°$ with the maximum of 1.36°.

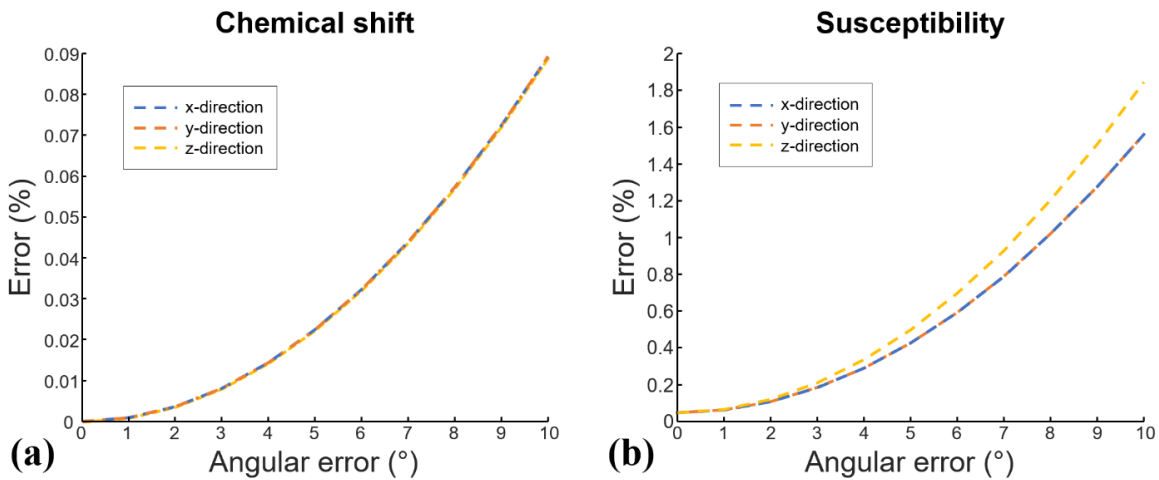

**Figure S3.** Evaluation of the misalignment effects of the cylindrical phantom. The plots of the percentage errors of the chemical shift/exchange (a) and susceptibility (b) measurements for the misalignment w.r.t. x- (blue), y- (orange), or z- (yellow) axis. The errors are small, suggesting the robustness of the method for the misalignment.

## Section 7. Effects of noise amplification

When considering noise ($n_x$, $n_y$, and $n_z$ with a standard deviation of $\sigma$ and uncorrelated to each other), Equation 6 becomes:

$$f_c(\vec{r}) = \frac{1}{3}\{f_x(\vec{r}) + n_x + f_y(\vec{r}) + n_y + f_z(\vec{r}) + n_z\}.$$

Hence, the noise component of the chemical shift/exchange map is $\frac{1}{3}(n_x + n_y + n_z)$ and the standard deviation is $\sqrt{E\left(\left(\frac{1}{3}(n_x + n_y + n_z)\right)^2\right)} = \frac{1}{\sqrt{3}}\sigma$. The final SNR of the chemical shift/exchange is increased by $\sqrt{3}$ due to the averaging. On the other hand, the noise effect in the susceptibility map is highly nonlinear and is dependent on the choice of a QSM reconstruction algorithm.